\documentclass[10pt,a4paper,twocolumn,english,prx,aps,showpacs,floatfix,groupedaddress,superscriptaddress]{revtex4-2}
\usepackage{graphicx,epsfig}
\usepackage{times}
\usepackage{graphics,dcolumn,bm}
\usepackage{amssymb,amsmath,rotate,color}
\usepackage{dsfont}
\usepackage[breaklinks,colorlinks=true,urlcolor=blue,citecolor=blue,linkcolor=blue]{hyperref}
\usepackage{soul}
\usepackage[normalem]{ulem}

\begin{document}
\newcommand{\be}{\begin{equation}}
\newcommand{\ee}{\end{equation}}
\newcommand{\bearr}{\begin{eqnarray}}
\newcommand{\eearr}{\end{eqnarray}}
\newcommand{\bseq}{\begin{subequations}}
\newcommand{\eseq}{\end{subequations}}
\newcommand{\nn}{\nonumber}
\newcommand{\dagg}{{\dagger}}
\newcommand{\vpdag}{{\vphantom{\dagger}}}
\newcommand{\bs}{\boldsymbol}
\newcommand{\up}{\uparrow}
\newcommand{\down}{\downarrow}
\newcommand{\fns}{\footnotesize}
\newcommand{\ns}{\normalsize}
\newcommand{\cdag}{c^{\dagger}}
\newcommand{\N}{_\text{N}}
\newcommand{\h}{_\text{H}}
\newcommand{\mg}{\Gamma_{\!\rm MG}}
\newcommand{\ctg}{\Gamma_{\!\rm CTG}}

\definecolor{red}{rgb}{1.0,0.0,0.0}
\definecolor{green}{rgb}{0.0,1.0,0.0}
\definecolor{blue}{rgb}{0.0,0.0,1.0}

\newcommand{\addFBA}[1]{\textcolor{blue}{#1}}
\newcommand{\commentFBA}[1]{\textcolor{blue}{{\bf FBA:} \emph{#1}}}

\title{Double-Exchange Enhanced Magnetic Blue-Shift of Mott Gaps}

\author{Mohsen Hafez-Torbati}
\email{mohsen.hafez@tu-dortmund.de}
\affiliation{Lehrstuhl f\"ur Theoretische Physik I, 
Technische Universit\"at Dortmund,
Otto-Hahn-Stra\ss e 4, 44221 Dortmund, Germany}
\author{Davide Bossini}
\email{davide.bossini@uni-konstanz.de}
\affiliation{Department of Physics and Center for Applied Photonics, University of Konstanz, Konstanz, Germany}
\author{Frithjof B. Anders}
\email{frithjof.anders@tu-dortmund.de}
\affiliation{Lehrstuhl f\"ur Theoretische Physik II, 
Technische Universit\"at Dortmund,
Otto-Hahn-Stra\ss e 4, 44221 Dortmund, Germany}
\author{G\"otz S. Uhrig}
\email{goetz.uhrig@tu-dortmund.de}
\affiliation{Lehrstuhl f\"ur Theoretische Physik I, 
Technische Universit\"at Dortmund,
Otto-Hahn-Stra\ss e 4, 44221 Dortmund, Germany}

\date{\today}%

\begin{abstract}
A substantial energy gap of charge excitations induced by 
strong correlations is the characteristic feature of Mott insulators.
We study how the Mott gap is affected by long-range antiferromagnetic order.
 Our key finding is that the Mott gap is increased by the magnetic
ordering: a magnetic blue-shift (MBS) occurs.
Thus, the effect is proportional to the exchange coupling in the leading order
in the Hubbard model. In systems with additional localized spins 
the double-exchange mechanism induces an  additional contribution to the MBS 
which is proportional to the hopping in the leading order. 
The coupling between spin and charge degrees of freedom bears the potential to enable spin-to-charge conversion 
in Mott systems on extreme time-scales determined by hopping and exchange only, 
since a spin-orbit mediated transfer of angular momentum is not involved in the process.
In view of spintronic and magnonic applications, it is highly promising to observe that several entire 
classes of compounds show exchange and double-exchange effects.
Exemplarily, we show that the magnetic contribution to the band-gap blue-shift observed
in the optical conductivity of $\alpha$-MnTe is correctly interpreted as the MBS of a Mott gap.
\end{abstract}

\maketitle

\section{Introduction}
\label{sec:intro}

The discovery of insulating behavior in  transition metal oxides \cite{Boer1937} and
 its explanation in terms of strong electron-electron interaction 
\cite{Mott1949} were the origin of the very active research field of 
strongly correlated systems \cite{Georges1996,Imada1998,Giannetti2016,Lee2006}.
The low-energy physics of Mott insulators is governed by spin excitations 
\cite{Manousakis1991}. A common and successful strategy to treat them consists in disentangling 
spin and charge degrees of freedom \cite{Fazekas99,Reischl2004,Hamerla2010,Hafez-Torbati2015}.
However, following this approach, it is difficult to track the coupling between charge and spin dynamics,
which is expected to play a pivotal role in the recent massive surge of interest in
antiferromagnetic (AF) spintronics 
\cite{Baltz2018,Sawatzky1984,Nemec2018,Kirilyuk2010,zaane85,Qaiumzadeh2018}. 
The grand goal of this impressive research effort consists in establishing the ability to convert 
spin signals into charge responses on the shortest possible timescale and minimizing 
as much as possible the energy dissipations. 

So far, the typical route to spintronics relies on spin-orbit based transport effects \cite{Baltz2018}
requiring a heavy metal layer on top of the antiferromagnet to read out the electric system. 
The vision of a spintronic information technology based solely on antiferromagnetism thus completely relies on the strength 
of the spin-orbit coupling, which defines both the spin-charge conversion efficiency and the operational frequency of a device.
Generically, exchange couplings are larger than spin-orbit couplings by at least one order of magnitude. 
This makes it desirable to use effects of purely exchange origin implying shorter characterstic time scales and hence 
higher operational frequencies.

Evidence has been reported that the charge gap in a Mott insulator, 
the Mott gap, depends on the magnetic ordering \cite{Sangiovanni2006,Wang2009,Fratino2017}. 
In particular, a magnetic shift of the band gap proportional to the 
square of the sublattice magnetization could enable a 
coherent modulation of the band gap energy itself.
Coherent dynamics of the order parameter in AF insulators has been photo-induced and manipulated 
\cite{Satoh2010,Kampfrath2011,Bossini2014} where frequencies of 22 THz \cite{Bossini2016,Hashimoto2018,Bossini2019}
were found. This framework would enable a coherent manipulation of the transport properties at the unprecedented 20 THz working frequency.

Local electronic interactions are essential for the formation of magnetic moments.
Thus, there are two possible dichotomous scenarios
for the influence of the magnetic ordering on the charge gap: 
(i) the charge gap is a band gap of $s$- and $p$-electrons which are \emph{different} from
the electrons forming the 
magnetic degrees of freedom. Then, the influence
of the localized spins is only indirect via superexchange with the itinerant
electrons. (ii) The charge gap is a Mott gap so that the electrons
forming the localized spins are also the ones forming the charge gap.
A charge-transfer insulator also belongs to scenario (ii)
because one of the bands relevant for the optical gap is a strongly
correlated one.

So far, the observed magnetic shifts have been discussed in terms of scenario
(i) \cite{Chou1974,Diouri1985,Ando1992}. The obtained results and even the overall sign of the 
effect, red- or blue-shift, depend on many details of the system. 
Only very recently, the observed magnetic shift of the band gap in hexagonal MnTe 
($\alpha$-MnTe) \cite{Bossini2020,Ferrer-Roca2000}
has been linked to strong local interactions in a local static mean-field model
\cite{Bossini2020}. 
In this article, we investigate the temperature dependence of the Mott gap 
across the transition from a paramagnetic to an AF insulator. 
The transition to the ordered phase 
is accompanied by a noticeable increase of the Mott gap, i.e.,
a magnetic blue-shift (MBS) of the Mott gap occurs. 

We address the fundamental nature of this MBS. By studying models with increasing complexity
within the dynamical mean field theory (DMFT) \cite{Georges1996} we are able to pinpoint the subtle differences and
clarify the influence of the charge hopping and the magnetic exchange interaction. We apply 
our approach to a real material, $\alpha$-MnTe, and demonstrate the very good agreement with the 
available experimental data.

First, we study the MBS in two fundamental models, the Hubbard model and the Hubbard-Kondo model, and 
unfold some generic features. In the Hubbard model, we show that in the leading order the
MBS is proportional to the magnetic exchange 
appearing by mapping the half-filled Hubbard model to a Heisenberg model or to a 
$t$-$J$ model (at finite doping). It results from the Slater mechanism:
as a static alternating field induced by the magnetic order opens an energy gap in a metal, such a field
also increases the existing charge gap in a Mott insulator.

In the Hubbard-Kondo model
an additional local spin $S$ is included at each lattice site to account for non-itinerant
magnetic degrees of freedom. Due to Hund's rules the itinerant and
the local spin are coupled locally by a ferromagnetic coupling $J_\text{H}>0$.
In this model, we find an additional contribution to the MBS which is
proportional to the hopping matrix element. 
This hopping contribution is the dominant contribution in systems with a large Hubbard interaction 
which implies a small magnetic exchange.
We reveal that this
contribution is induced by the double-exchange mechanism due to a reduced effective hopping upon 
transition from the paramagnetic to the AF insulator.

Second, we verify our approach by analyzing an exemplary system promising for applications, 
hexagonal MnTe ($\alpha$-MnTe), which has been experimentally investigated \cite{Ferrer-Roca2000,Bossini2020}.
We extend the $S=5/2$ spin model explaining the inelastic neutron 
scattering data \cite{Szuszkiewicz2006,Mu2019} to a
Hubbard-Kondo model allowing for the coupling of the spin and
 charge degrees of freedom.
We compute the MBS of the Mott gap of the half-filled $3d$-shell 
of Mn ions. Using only generic parameters established for $\alpha$-MnTe in literature and 
without any fine-tuning we achieve an overall excellent description of the MBS measured in the 
optical conductivity. We unveil the origin of the MBS in $\alpha$-MnTe data and find a magnetic 
exchange contribution of $36\%$ and a hopping contribution of $64\%$.
Our findings set the stage to study coupled spin and charge dynamics 
in strongly correlated systems, including the specific case of $\alpha$-MnTe.

The article is organized as follows. After this Introduction, results for the
Hubbard model are shown and interpreted. Subsequently, the results for the 
Hubbard-Kondo model are presented and discussed, in particular the additional
contribution to the MBS stemming from the double-exchange mechanism.
Section \ref{sec:MnTe} deals with the particular case of $\alpha$-MnTe
as a candidate for significant spin-charge coupling based on the MBS.
In Section \ref{sec:conclusio} the results are summarized and a brief outlook is given.

\section{The three-dimensional Hubbard model}
\label{sec:3DH}

The Hubbard model \cite{Hubbard1963} at half-filling comprises hopping 
between nearest-neighbor (NN) sites 
controlled by the parameter $t$ and an interaction $U$ between electrons at the same
site with opposite spins 
\be
\label{eq:HM}
H_{\rm H}=-t\!\!\sum_{<i,j>}\sum_{\sigma=\uparrow,\downarrow} 
\left( c^{\dagg}_{j,\sigma}c^{\vpdag}_{i,\sigma} 
+ {\rm H.c.} \right)+ U \sum_{i} n^{\vpdag}_{i,\downarrow} n^{\vpdag}_{i,\uparrow}  .
\ee
This well studied model shows a particle-hole symmetry at half-filling with respect to the energy $\mu=U/2$ defining 
the chemical potential $\mu$ used throughout this paper.
The phase diagram at finite temperatures on the cubic lattice
is well known at half-filling \cite{Rohringer2018,Kent2005,Staudt2000,Ulmke1995,Jarrell1992}. 
At $T=0$, the ground state is a N\'eel antiferromagnet for any finite $U/t$ \cite{Pruschke2003}.
The N\'eel temperature $T_{\rm N}$ separating the paramagnetic and  the AF phases increases from 0 upon increasing $U$ \cite{Staudt2000}, 
reaches a maximum and decreases as $T_{\rm N}\propto t^2/U$ in the strong coupling limit $U\gg t$ where the model can be mapped onto a spin-1/2 Heisenberg model.
At high temperatures, the phase is a metal for small $U/t$  and a paramagnetic Mott insulator for large $U/t$ separated from the metallic phase by a 
crossover region.
At large $U/t$, the Mott gap is proportional to $U$ as a charge 
excitation leads to the creation of a double occupancy
which requires the energy $U$.

We define the bare charge gap $\Delta$ as the value of the charge gap in the absence of the hopping, $t=0$. 
Although in the Hubbard model the bare charge gap just equals the Hubbard interaction $U$, it remains 
the relevant quantity also in the Hubbard-Kondo model defining the magnetic exchange interaction $J=4t^2/\Delta$.
We use the bare charge gap $\Delta$ in cases we aim to compare our results for the Hubbard model and the 
Hubbard-Kondo model.

We employ the DMFT \cite{Georges1996} with exact diagonalization (ED) \cite{Caffarel1994}
as impurity solver. 
This approach is well established for strong local interactions where subtle effects 
such as an emerging exponentially low-energy scale associated with the formation of a narrow band 
at the chemical potential in the metallic phase cannot occur. For more details on the method we refer to  Appendix \ref{app:A}.
We compute the 
averaged local spectral function $A(\omega)= (A^A_\sigma(\omega)+A^B_\sigma(\omega))/2$
 from the imaginary part of the 
local Green function of sublattice $A$ and $B$. We point out that the spectral function $A(\omega)$ does not depend on spin 
even in the AF phase because we average over both sublattices.

\begin{figure}[t]
   \begin{center}
 \includegraphics[width=0.9\columnwidth,angle=-90]{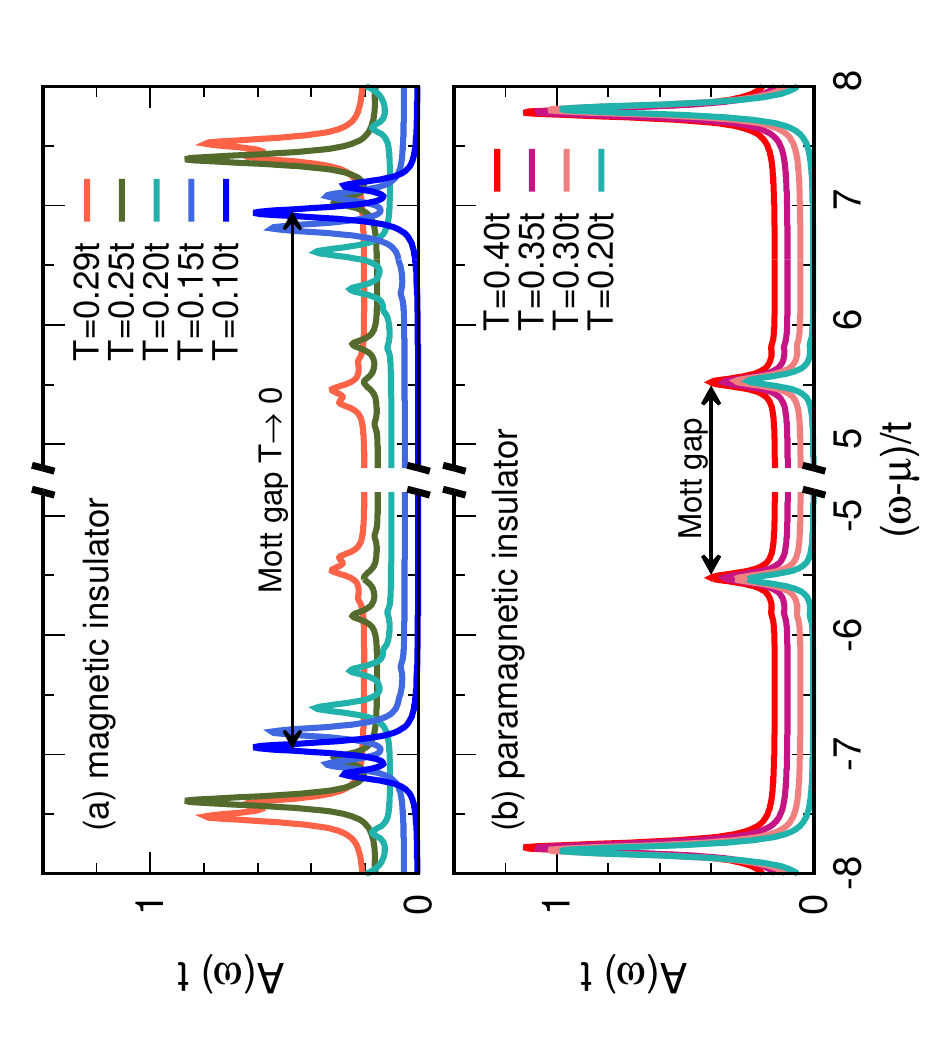}
   \caption{Spectral function $A(\omega)$ vs.\  $\omega$ 
   in the range $[-8t,+8t]$ at various temperatures $T$
   in the magnetic insulator (MI) (a) and in the paramagnetic insulator (PI) 
	(b) for $U=20t$ and $n_b=6$ bath sites in the impurity problem.
	The N\'eel temperature is given by $T_{\rm N}\approx 0.3t$.}
   \label{fig:sp:hm:T}
   \end{center}
\end{figure}

\begin{figure}[b]
   \begin{center}
 \includegraphics[width=0.45\columnwidth,angle=-90]{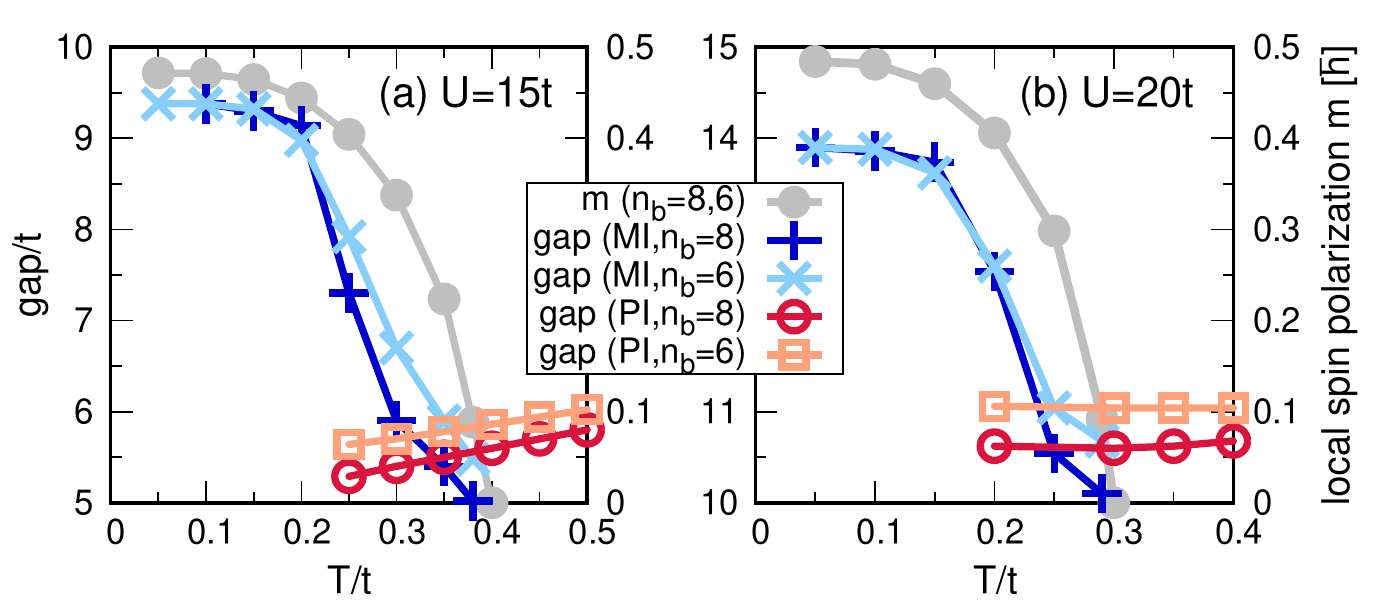}
   \caption{The Mott gap in the magnetic insulator (MI) and in the paramagnetic insulator (PI) 
	as function of temperature for  $U=15t$ (a) and $U=20t$ (b). 
	The grey lines show the local spin polarization $m$ (right axes). 
	The results for $n_b=6$ and $n_b=8$ are compared.}
   \label{fig:mg:hm}
   \end{center}
\end{figure}

We plot the local spectral function $A(\omega)$ of the Hubbard model 
for different temperatures  in the magnetic insulator (MI)
phase in Fig.\ \ref{fig:sp:hm:T}(a)
and in the paramagnetic insulator (PI) phase in Fig.\ \ref{fig:sp:hm:T}(b)
for $U=20t$.
The spectral functions for the different parameters are shifted vertically for clarity. 
Note that the peak structure in $A(\omega)$ is caused by the discretized 
representation of the conduction band in the ED impurity solver with the number 
of bath sites $n_b=6$.
Lowering the temperature $T$ in the PI hardly changes $A(\omega)$. 
But in the MI, a shift of the electron 
and hole contributions to higher excitation energies is clearly observed. 
Below the N\'eel temperature $T_{\rm N}\approx 0.3t$, the stable phase
is the MI, but the metastable PI solution can be computed as well and 
was added to Fig.\ \ref{fig:sp:hm:T}(b) for comparison.

The Mott gap is obtained from the energy difference between the two excitation 
energies of the spectrum that are closest to the chemical potential $\mu$, see the 
indicated arrows in Fig.\ \ref{fig:sp:hm:T}.
While this gap is apparently a constant in the PI phase as can be seen in 
Fig.\ \ref{fig:sp:hm:T}(b), it shows a strong temperature dependency in the MI phase 
in Fig.\ \ref{fig:sp:hm:T}(a). Upon reducing $T$, the electron and hole 
peaks at $\pm 5.5t$ shift apart to $\pm 7t$ due to the magnetic ordering. 
This leads to a MBS of the Mott gap $\mg(T)$
of about $3t$ as $T\to 0$.

We depict the Mott gap as well as the  local spin polarization $m$
in units of $\hbar$  vs.\ the temperature $T$ 
for $U=15t$ in Fig.\ \ref{fig:mg:hm}(a)
and for $U=20t$ in Fig.\ \ref{fig:mg:hm}(b). The
results are displayed for two bath sizes, $n_b=6$ and $n_b=8$
to illustrate the accuracy of the approach.
For $U=15t$, the gap in the PI decreases slowly upon lowering the
temperature. For $U=20t$ it remains almost constant. 
However, in both cases there is a rapid increase of the gap upon entering
the MI phase which illustrates the MBS.

At the continuous transition from the PI to the MI the gaps have to be equal.
This is not quite the case, most likely because
of inaccuracies in extracting the gap  from the ED data
at finite bath sites.
For $U=15t$, the gap value is about $5.5t$ close to
the transition temperature and rises to about $9.4t$ for $T\to 0$.
Comparing Figs.\ \ref{fig:mg:hm}(a) and \ref{fig:mg:hm}(b) a decrease
of the MBS upon increasing $U$ from $15t$ to $20t$ is observed. Such a decrease of 
the MBS in the Mott regime has also been observed in Ref.\ \cite{Fratino2017}.

\begin{figure}[t]
   \begin{center}
   \includegraphics[width=0.74\columnwidth,angle=-90]{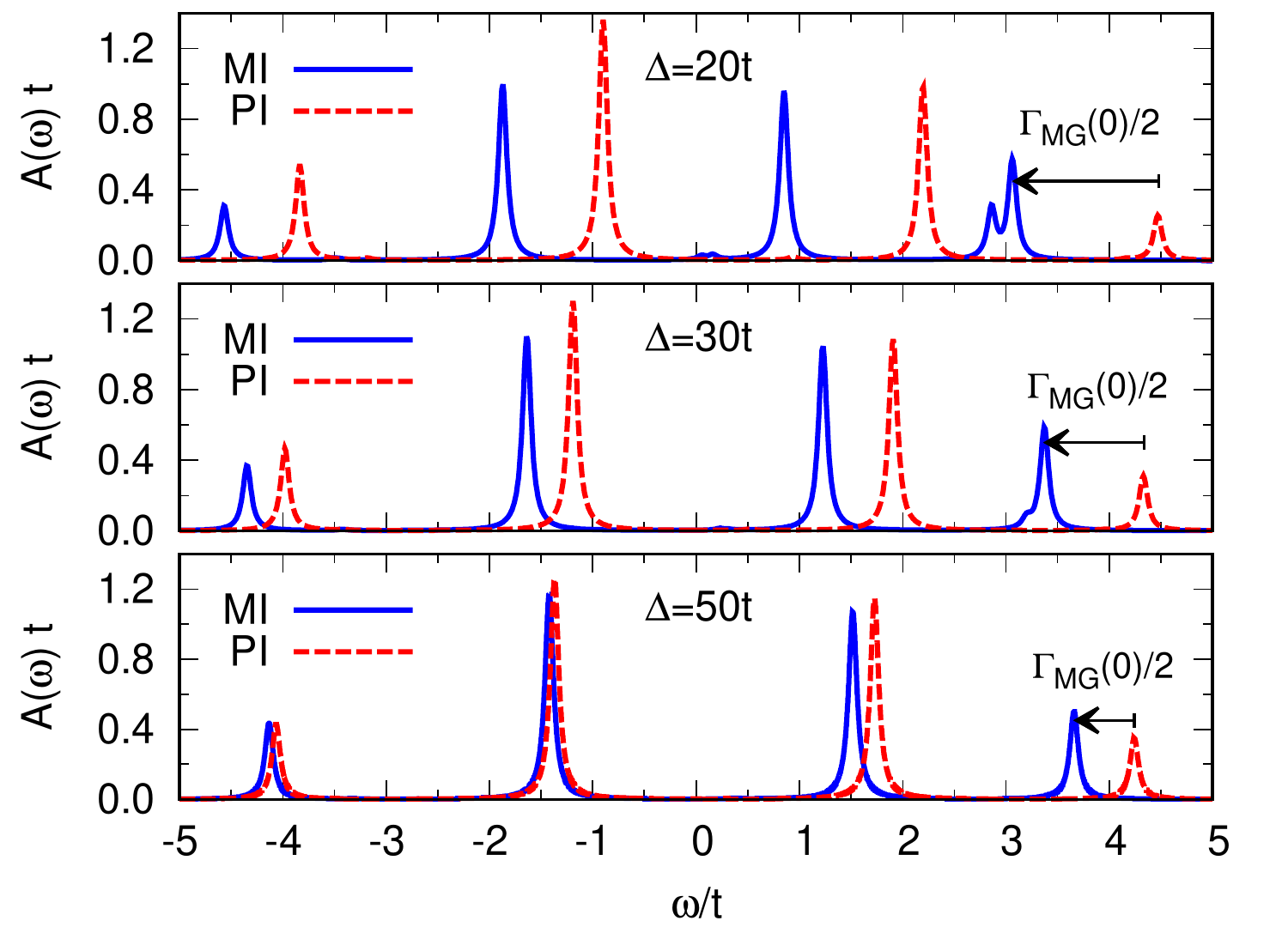}
   \caption{The lower band of the spectral function in the Hubbard model for 
	 various values of the bare charge gap $\Delta=U$ in the MI
   close to $T=0$ and in the PI close to  $T=T_{\rm N}$. The Fermi energy is located at $\omega=\mu=U/2$. 
   We have used the bare charge gap $\Delta$ as
   the label since we aim to compare the results with the results of the Hubbard-Kondo model. Clearly, the   
	 spectral functions in the PI and in the MI phase approach each other upon increasing $\Delta$.
	 The results are obtained for $n_b=6$ bath sites in the impurity solver. 
	 The MBS of the Mott gap $\mg(0)$ is twice the indicated arrows.}
   \label{fig:sp_hm_T0}
   \end{center}
\end{figure}

To analyze the MBS $\mg(T)$ near $T=0$ further,
we depict the spectral function as function of $\omega$ for various values of the bare charge gap 
$\Delta=U$ in the MI close to $T=0$ and in the PI close to $T=T_{\rm N}$
in Fig.\ \ref{fig:sp_hm_T0}. We have used the bare charge gap $\Delta$ as the label since we aim to compare
the results with the results of the Hubbard-Kondo model in the next section. 
Only the lower Hubbard band is shown in Fig.\ \ref{fig:sp_hm_T0} because 
the upper Hubbard band is its mirror image with respect to $\omega=\mu=U/2$ due to electron-hole symmetry, 
which is perfectly realized in our numerical data.
We indicated the shift of the excitation peak closest to the chemical potential by an arrow
defining half of the MBS, $\mg(0)/2$. The results are for $n_b=6$ bath sites in the impurity solver.
Fig.\ \ref{fig:sp_hm_T0} reveals that the spectral functions in the MI and in the PI approach each other more and more upon 
increasing the bare charge gap $\Delta=U$. Consequently, the MBS $\mg(0)$ decreases as $t/\Delta\to 0$.

The MBS in the Hubbard model has already been observed in previous work \cite{Sangiovanni2006,Wang2009} 
but not systematically studied. Recently, its monotonic decrease in the Mott regime for $U\rightarrow \infty$ was noted in 
Ref.\ \cite{Fratino2017}. But so far neither a functional dependence nor a physical interpretation has been given.
The microscopic understanding of this highly promising effect for application in AF spintronics is thus still lacking.

\begin{figure}[t]
   \begin{center}
   \includegraphics[height=0.99\columnwidth,angle=-90]{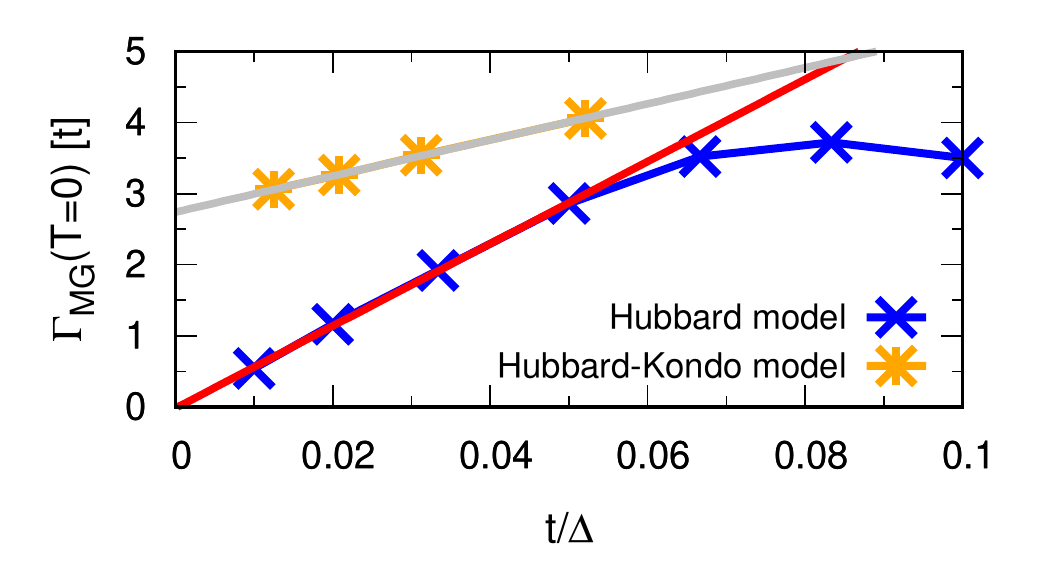}	
   \caption{The magnetic blue-shift of the Mott gap $\mg(T)$ in units of the hopping $t$ 
	at $T=0$  is plotted vs.\ $t/\Delta$, where $\Delta$ 
  is the bare charge gap. The results are obtained 
	for $n_b=6$ bath sites in the impurity problem. In the Hubbard-Kondo model
	we set the local spin to $S=2$ and the Hund coupling to $J_\text{H}=0.15U$.}
   \label{fig:mbs}
   \end{center}
\end{figure}

In order to provide a quantitative description
of the influence of the hopping $t$ and the bare charge gap $\Delta$ onto the MBS
we plot $\mg(0)/t$ vs.\ $t/\Delta$ in Fig.\ \ref{fig:mbs}
for various combinations of $t$ and $\Delta$. This demonstrates clearly
that the MBS in the Mott regime is proportional to
the exchange coupling $J=4t^2/\Delta$. Since the figure renders
$\mg(0)$ in units of $t$ the proportionality 
$\mg(0) \propto J$ implies a straight line as depicted in red.
It fits very well to the blue data for small values of $t/\Delta$ with a slope of $57.8$
which is equivalent to $\mg(0) \approx 14.4 J$ underlining its magnetic origin.
Thus, this effect is quite sizable and sets the scale for further contributions.

The MBS can also be linked to the  decrease of the free energy when the 
system enters the MI phase. If such a decrease did not occur
the system would not display the phase transition to the ordered phase.
The free-energy change below $T_{\rm N}$ is mainly due to the reduction of 
the internal energy, which can be determined solely from the single-particle spectral 
function. Upon transition from the PI to the MI a redistribution of the weight within the 
spectral function occurs which leads to a large \emph{increase} in the internal energy if it were not 
compensated by a MBS. This is exemplarily illustrated in App.\ \ref{app:D}
and corroborates that the MBS is a generic feature upon entering an antiferromagnetically
ordered phase.

\section{Hubbard-Kondo model}
\label{sec:HKM}

We extend the analysis presented so far for the Hubbard model to a model which includes
localized spins so that it also bears features of a Kondo system. Specifically, we consider the 
Hubbard-Kondo model \cite{Doniach1977,Fulde1993,Held2000} given by
\bseq
\begin{align}
H_\text{HK} &= H_{\rm H} + H_\text{K}
\\
H_{\rm K} &=-2J_{\rm H}\sum_{i}\vec{s}_i \cdot \vec{S}_i \ ,
\end{align}
\eseq
where $H_{\rm H}$ is the Hubbard model Eq. \eqref{eq:HM}. The Kondo term $H_\text{K}$ couples the spin of the 
electron $\vec{s}_i$ ferromagnetically to the local spin $\vec{S}_i$
originating from a Hund's coupling.
We choose the local spin quantum number to be $S=2$ and  
the Hund coupling to be $J_{\rm H}=0.15U$. The bare charge gap is no longer given by
$U$ alone but acquires a contribution from the Hund's coupling, $\Delta=U+4J_{\rm H}$.

\begin{figure}[t]
   \begin{center}
   \includegraphics[width=0.74\columnwidth,angle=-90]{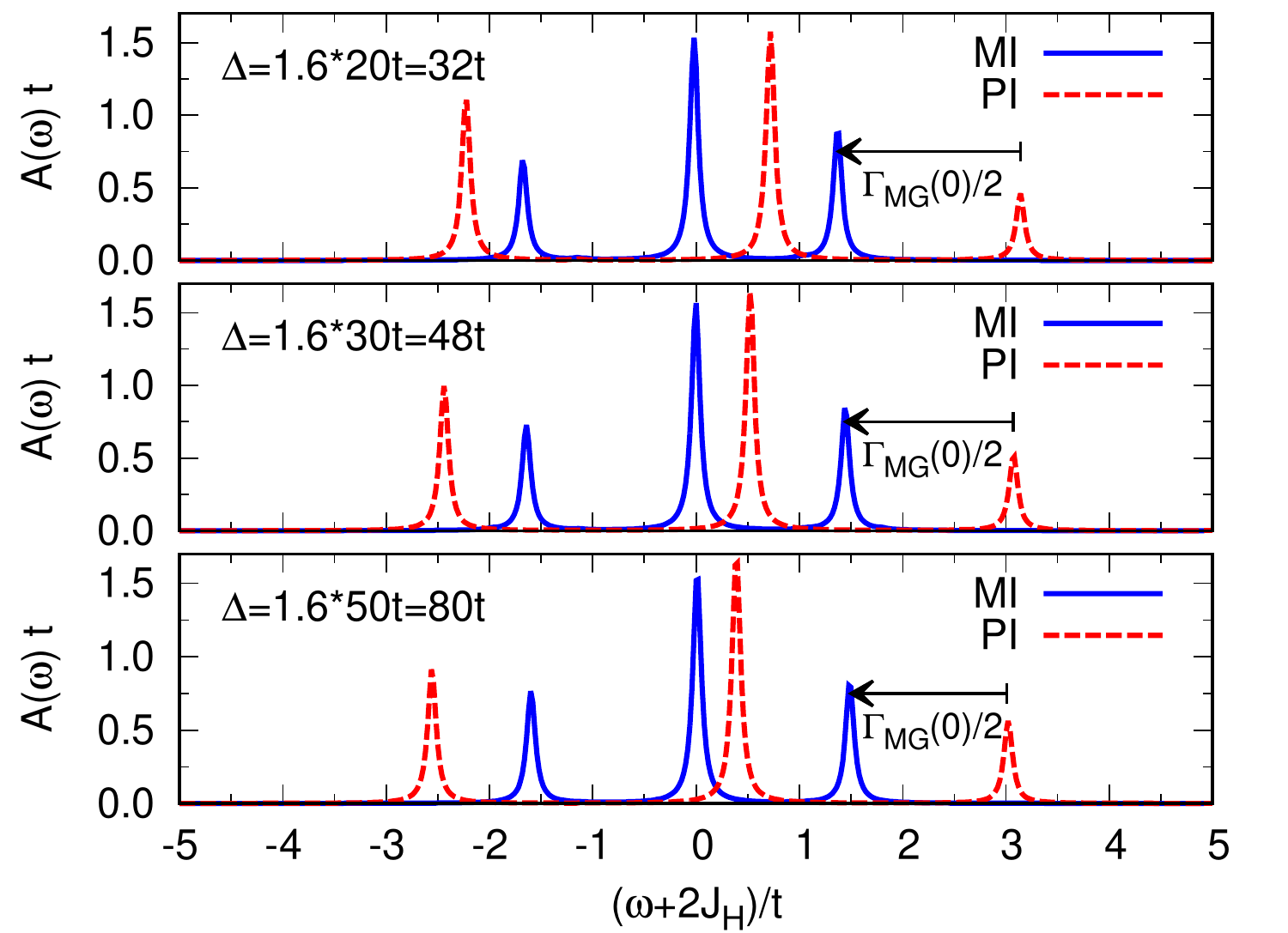}
   \caption{The same as Fig. \ref{fig:sp_hm_T0}, but for the Hubbard-Kondo model 
   with the local spin $S=2$ and the Hund coupling $J_{\rm H}=0.15U$. 
	The bare charge gap is given by $\Delta=U+4J_{\rm H}=1.6U$.
	The spectral functions do not approach each other upon increasing $U$.}
   \label{fig:sp_hkm_T0}
   \end{center}
\end{figure}

Fig.\ \ref{fig:sp_hkm_T0} depicts the local spectral functions as in Fig.\ \ref{fig:sp_hm_T0} but
for the Hubbard-Kondo model. 
In contrast to data from the Hubbard model, Fig.\ \ref{fig:sp_hkm_T0} shows that the
spectral functions in the PI and in the MI remain distinctly different even for large $\Delta$
resulting in an enhanced MBS of the Mott gap $\mg(0)$, see the indicated arrows. 
This can clearly be associated to the noticeably smaller bandwidth in the MI phase compared to 
the PI phase, which is the fingerprint of the double-exchange mechanism \cite{Zener1951,Anderson1955,Gennes1960,Pavarini2012}.

\begin{figure}[b]
   \includegraphics[width=0.9\columnwidth]{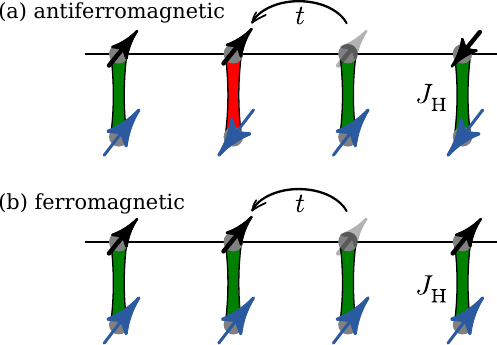}
    \caption{Illustration of the reduction of the effective hopping
		due to the double-exchange mechanism. Due to the strong Hund coupling
		only electrons with spin aligned with the local spin can occur. This 
		allows unrestricted hopping between sites with parallel spin orientation,
		see panel (b). 
		But for antiparallel spin orientation 
		no hopping is possible, see panel (a).}
   \label{fig:DX}
\end{figure}

The double-exchange mechanism is well known for enhancing the mobility of an electron 
in a ferromagnetic state since the 1950s,
and it is responsible for ferromagnetism in perovskite manganites. 
In an AF state, however, the double-exchange mechanism strongly suppresses the effective hopping between 
sites with antiparallel spin ordering.

The basic idea of the mechanism  is illustrated in Fig. \ref{fig:DX}. A hole added to the half-filled system
propagates with an effective hopping which determines the bandwidth of the single-particle spectral function. 
In the limit of large $J_{\rm H}$ it is natural to restrict the local Hilbert space 
such that the electron spin $s=1/2$ and the local spin $S$ always form the maximum total spin $S+1/2$. 
This allows to derive the following relation for the effective hopping between sites $i$ and $j$ \cite{Mueller-Hartmann1996}
\be
\frac{t^{\rm eff}_{i,j}}{t}=\frac{\mathcal{S}^{T}_{i,j}+1/2}{2S+1}(-1)^{2S-\mathcal{S}^{T}_{i,j}+1/2} \ ,
\label{eq:eh}
\ee
where $\mathcal{S}^{T}_{i,j}$ is the total {\it bond} spin construced from the spin at site $i$ 
and the spin at site $j$, i.e., from $S+1/2$ and $S$ \footnote{One notes that we have denoted the local spin by $S$ while in Ref. \cite{Mueller-Hartmann1996} 
it is denoted by $S-1/2$.}. 
One notes that there is always a hole 
either at site $i$ or at site $j$. The total bond spin takes the values 
$\mathcal{S}^{T}_{i,j}=1/2, 3/2, \cdots, 2S+1/2$. For parallel spin ordering between sites 
$i$ and $j$ we have $\mathcal{S}^{T}_{i,j}=2S+1/2$ which results in the effective hopping $t^{\rm eff}_{i,j}=t$. 
The PI phase is described by singlet bonds, i.e., bonds with the total spin $0$. Adding a hole to a singlet bond 
creates a bond with the total spin $\mathcal{S}^{T}_{i,j}=1/2$, which can be realized from the commutation
relation
\be
\left[\vec{\mathcal{S}}^{\ T}_{i,j} \cdot \vec{\mathcal{S}}^{\ T}_{i,j}, c^{\vpdag}_{j,\alpha}  \right]=
\frac{3}{4}c^{\vpdag}_{j,\alpha}-\sum_{\beta}c^{\vpdag}_{j,\beta} 
\vec{\sigma}^\vpdag_{\alpha,\beta}\cdot \vec{\mathcal{S}}^{\ T}_{i,j} \quad ,
\ee
where $\vec{\mathcal{S}}^{\ T}_{i,j}=\vec{s}_i + \vec{S}_i+ \vec{s}_j + \vec{S}_j$ is the total bond spin 
operator, and $\vec{\sigma}$ is a vector made of Pauli matrices. Such a hole propagates with the 
effective hopping $t^{\rm eff}_{i,j}=t(-1)^{2S}/(2S+1)$ according to Eq. \eqref{eq:eh}. 
In the case of antiparallel spin ordering between sites $i$ and $j$, a pure hopping can never take place, i.e.,
the hopping of the hole is always accompanied by the reduction of the local magnetic numbers 
from the absolute maximum values \cite{Mueller-Hartmann1996}.

The above discussion explains the narrower bandwidth we observe for the Hubbard-Kondo model 
in Fig. \ref{fig:sp_hkm_T0} in contrast to the results for the Hubbard model in Fig. \ref{fig:sp_hm_T0}.
One notes that the effective hopping in the Hubbard model for both MI and PI phases is the 
bare hopping $t$.
More importantly, the above discussion explains the narrower bandwidth we observe in the MI phase 
in contrast to the PI phase in Fig. \ref{fig:sp_hkm_T0}, which is the origin of the enhanced 
MBS in the Hubbard-Kondo model.

The results for the MBS in the Hubbard-Kondo model at $T=0$ in units of $t$ are included in Fig. \ref{fig:mbs}.
The qualitative behavior of the MBS in the Hubbard-Kondo model significantly differs from 
those of the Hubbard model. A substantial offset in the limit $t/\Delta \to 0$ is observed 
in the quantity  $\mg(0)/t$. A linear fit given by the gray line
\be
\label{eq:fit-HK-graph}
\frac{\Gamma_{\rm MG}(0)}{t} =C_1 +4 C_2 \frac{t}{\Delta}
\ee 
with the constants $C_1=2.7$ and $C_2=6.3$ nicely agrees with our data.
Note that the exchange coupling is given by $J=4t^2/\Delta$ such that
we end with the fit
\be
\label{eq:fit-HK}
\Gamma_{\rm MG}(0)=C_1 t +C_2 J \quad.
\ee 
By plotting $\mg(0)/t$ vs.\ $t/\Delta$ in Fig.\ \ref{fig:mbs} we can separate the two different 
contributions to the MBS more clearly: one proportional to the hopping $t$ which appears as a 
constant term and one proportional to the magnetic exchange $J$ which appears as a linear term. 
The first contribution results from changes in the effective hopping and we refer to it as the hopping 
or the double-exchange contribution. The second contribution results
from the alternating magnetic field as in the Hubbard model and we refer 
to it as the exchange contribution.

Our findings unfold the essential role that the double-exchange mechanism 
can play in the future development of AF spintronic: It induces a coupling 
between the magnetic order and the charge gap as large as the hopping.
The relation Eq. \eqref{eq:fit-HK} is highly promising since there are several entire classes
of compounds which show the exchange and the double-exchange effects. 
We leave a more detailed investigation of the hopping and the exchange contributions 
of the MBS to future research, and instead apply our approach to a real material for the 
rest of this paper.

\section{Application to $\alpha$-M\lowercase{n}T\lowercase{e}}
\label{sec:MnTe}

Now we apply the acquired understanding of the MBS in the Hubbard-Kondo model
to a real compound: $\alpha$-MnTe.
This AF semiconductor displays a noticeable additional increase of the optical gap below
its N\'eel temperature $T\N\approx 310$ K. To separate the MBS 
from other temperature dependent contributions, which are continuous, the experimental band gap is fitted
in the paramagnetic regime $T>T\N$ by the empirical Varshni function \cite{Varshni1967}
which allows one to extrapolate the temperature dependence 
of the band gap in a paramagnetic semiconductor down
to zero temperature. The difference of the actually measured 
gap to the extrapolated value
quantifies the MBS \cite{Ferrer-Roca2000,Bossini2020}.
Similar analyses were performed also for other magnetic semiconductors
\cite{Chou1974,Diouri1985,Ando1992,Zhu2018}.

\begin{figure}[t]
   \begin{center}
   \includegraphics[width=0.99\columnwidth]{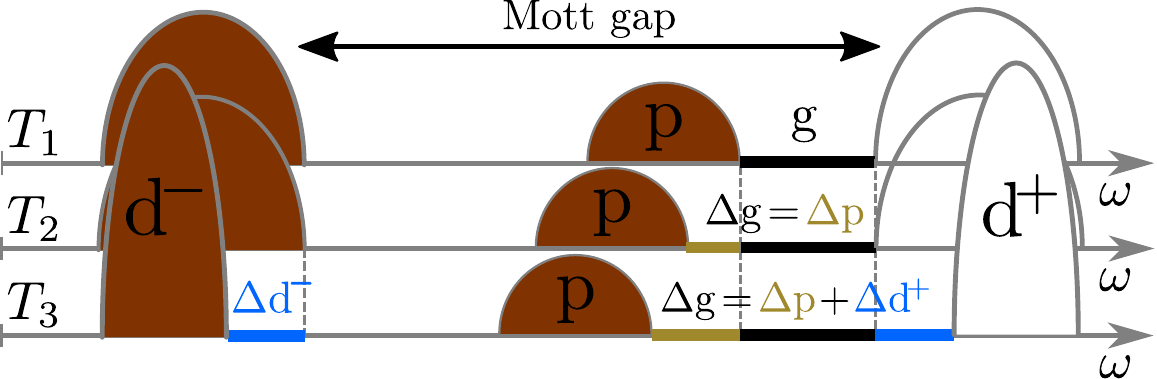}
   \caption{Sketch of the bands in $\alpha$-MnTe 
	comprising the $p$-bands at Te$^{2-}$and the lower ($d^-$) and upper ($d^+$)
	Hubbard bands of the $3d$-electrons at Mn$^{2+}$
   at three temperatures $T_1>T_2>T\N>T_3$, 
   where $T\N$ is the N\'eel temperature. The energy difference between the $d^-$ and the $d^+$ Hubbard bands 
	 defines the Mott gap and the energy difference between the $p$- and the $d^+$-band defines the charge-transfer gap $g$. 
	 The optical gap equals the charge-transfer gap. The Mott gap experiences the magnetic blue-shift 
	 $\Gamma_{\rm MG}(T_3)=\Delta d^+ + \Delta d^-$ while the charge-transfer gap experiences the magnetic blue-shift 
	 $\Gamma_{\rm CTG}(T_3)=\Delta d^+$, see main text.}
   \label{fig:bandgap}
   \end{center}
\end{figure}

The magnetic order in $\alpha$-MnTe consists of planes of Mn$^{2+}$ ions forming triangular lattices 
in which spins are parallelly ordered. These planes are stacked
and the spins are oriented antiparallel in adjacent planes generating AF order.
According to Hund's rule the total spin at the Mn$^{2+}$ ions is $S=5/2$
due to the half-filled $d$-shell. The dispersion of the collective magnons 
is well understood \cite{Szuszkiewicz2005,Szuszkiewicz2006,Mu2019}.
In contrast, the knowledge of the electronic excitations is significantly 
less developed, and the understanding of its coupling to the magnetic system is still 
in its infancy. Density-functional calculations (DFT) \cite{Youn2004,yin19,Mu2019} 
indicate that the conduction band 
in $\alpha$-MnTe is dominated by Mn $3d$-contributions 
although the Mn $4s$-orbital is also involved. 
Assuming scenario (ii) we neglect the $4s$-admixture and treat $\alpha$-MnTe as a 
charge-transfer insulator \cite{zaane85} where
the optical gap $g$ arises from promoting an electron from the filled 
$p$-band of Te$^{2-}$ to the empty upper Hubbard band $d^+$ at Mn$^{2+}$; cf.\
the panel for $T_1$ in Fig.\ \ref{fig:bandgap}.

Fig.\ \ref{fig:bandgap} schematically depicts the relative change
of bands upon lowering the temperature $T_1 \to T_2 \to T_3$. 
As the temperature is decreased from $T_1$ to $T_2>T\N$ the $p$- and the $d^+$-bands 
shift apart with the Mott gap remaining unchanged.
This increase of the charge-transfer gap $g$ is induced by slight structural
changes and minute temperature effects in the
paramagnetic phase. This fraction of the change of the gap is 
continuous through the magnetic transition
and thus captured by the extrapolation with the Varshni function.
For $T_3<T\N$ an additional contribution $\Delta d^+$ to the gap
arises  due to the MBS of the upper Hubbard band $d^+$. 
In principle, the magnetic ordering could affect also the $p$-band, 
but this would be an indirect effect and we thus assume it to be less relevant. 
Consequently, to address the MBS in $\alpha$-MnTe we focus on an
effective Hamiltonian describing the electrons in the $3d$-shell of the Mn$^{2+}$ ions.

For a quantitative description, the
established Heisenberg model for the spins of the Mn$^{2+}$ ions
\cite{Szuszkiewicz2006,Mu2019} needs to be extended by the charge 
degrees of freedom. The full extension would require to consider at least
five $d$-bands from the Mn$^{2+}$ ions plus three $p$-bands from the Te$^{2-}$ ions. 
This is by far too complex for an explicit
numerical treatment of the strong interactions present at the Mn-sites.
For this reason, we follow the idea proposed in Ref.\  \cite{Bossini2020}
and describe the itineracy of each of the five $d$ electrons
in a one-band Hubbard model while treating the other four $d$-electrons
as localized forming a spin $S=2$. We stress that the itinerant electron 
is a representative for all five electrons. We do not claim
that the five orbitals are different, but that for each electron
in one of them the other four act like a localized spin. 
In other words, we make the approximation that the local Fock space of 
the Mn$^{+2}$ $3d$-orbitals is restricted to the charge configurations $N=4$, 5, and 6, 
so that we only need to take into account the charge fluctuation in one effective 
local orbital which is degenerate with respect to spin. This is well justified since we are interested in 
the low-energy charge excitations, specifically, the charge gap.

\begin{figure}[t]
   \begin{center}
   \includegraphics[width=\columnwidth,angle=0]{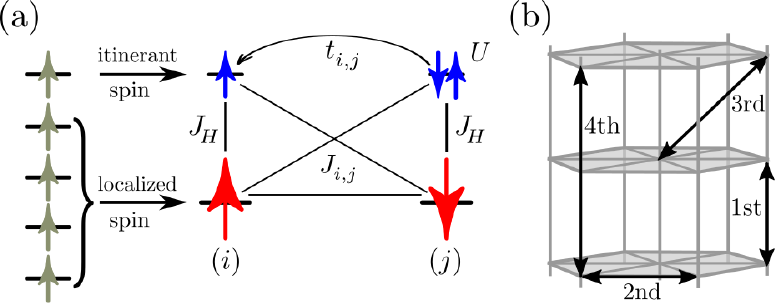}
   \caption{(a) Illustration of the Hubbard-Kondo model 
	 \eqref{eq:HKM_main} for the half-filled $3d$-shell 
   of Mn$^{2+}$-ions at two sites $i$ and $j$. 
   (b) Stacked triangular layers with 1st, 2nd, 3rd, and 4th 
   neighbor specified so that we distinguish $t_1, t_2, t_3, t_4$ and
	$J_1, J_2, J_3, J_4$.}
   \label{fig:HKM}
   \end{center}
\end{figure}

Hence, we consider a Hubbard-Kondo lattice model on stacked
triangular lattices, cf.\ Fig.\ \ref{fig:HKM}, 
\bearr
\label{eq:HKM_main}
&&H=-\sum_{i,j}\sum_{\sigma=\uparrow,\downarrow} t_{i,j}^\vpdag 
( c^{\dagg}_{j,\sigma}c^{\vpdag}_{i,\sigma}+{\rm H.c.}) 
+ U \sum_{i} n^{\vpdag}_{i,\downarrow} n^{\vpdag}_{i,\uparrow} \nn 
\\ 
&\!-\!& 2J^\vpdag\h\sum_{i} \vec{S}_i \cdot \vec{s}_i
+\sum_{i,j} J_{i,j} ( \vec{S}_i \cdot \vec{s}_j +
\vec{S}_j \cdot \vec{s}_i +\vec{S}_i \cdot \vec{S}_j)
\eearr
where the $t_{i,j}$ are the hopping elements and 
the $J_{i,j}$ the magnetic couplings, see Fig.\ \ref{fig:HKM}(a).
These effective magnetic couplings $J_{i,j}$ 
result from virtual excitations of the four 
$d$-orbitals, that are treated as local, to the neighboring Mn sites.

The intersite couplings 
are limited to the four nearest neighbors specified in Fig.\ \ref{fig:HKM}(b). We 
denote the hopping and the magnetic coupling of $n$th neighbor by $t_n$ and $J_n$.
The magnetic couplings are taken from the
measured magnon dispersion \cite{Mu2019} to be
$J_1=3.072$ meV, $J_2=0.0272$ meV, $J_3=0.4$ meV, and $J_4=0.16$ meV,
matching also the observed N\'eel temperature. 
The Hubbard interaction $U$ ranges between $\approx 5$ eV
to $\approx 7$ eV and the Hund coupling between $\approx 0.7$ eV to $\approx 1.0$ eV,
based on estimates from atomic physics \cite{Bossini2020} and 
the DFT \cite{Youn2004,Mu2019} calculations.
We investigate the effect of $U$ and $J\h$ on the MBS in this  parameter regime.
The hopping elements $t_n$ are determined such that they are consistent with
the intersite exchange couplings, i.e., $J_n = 4t^2_n/\Delta$ where $\Delta$ is the 
bare charge gap $U+4J\h$. This is to guarantee that the low-energy spin excitations 
of the Hamiltonian Eq. \eqref{eq:HKM_main} is described by the $S=5/2$ Heisenberg model 
already established for $\alpha$-MnTe by inelastic neutron scattering measurements \cite{Mu2019}.
The explicit values of the parameters are given in App.\ \ref{app:E}.
It must be noted that the larger, dominant hoppings $t_1$ and $t_3$ link sites 
with AF ordering.  Hence we expect a noticeable hopping contribution to the MBS 
to occur, stemming from the double-exchange mechanism described in Sec.\ \ref{sec:HKM}.

\begin{figure*}[t]
   \begin{center}
   \includegraphics[width=0.65\columnwidth,angle=-90]{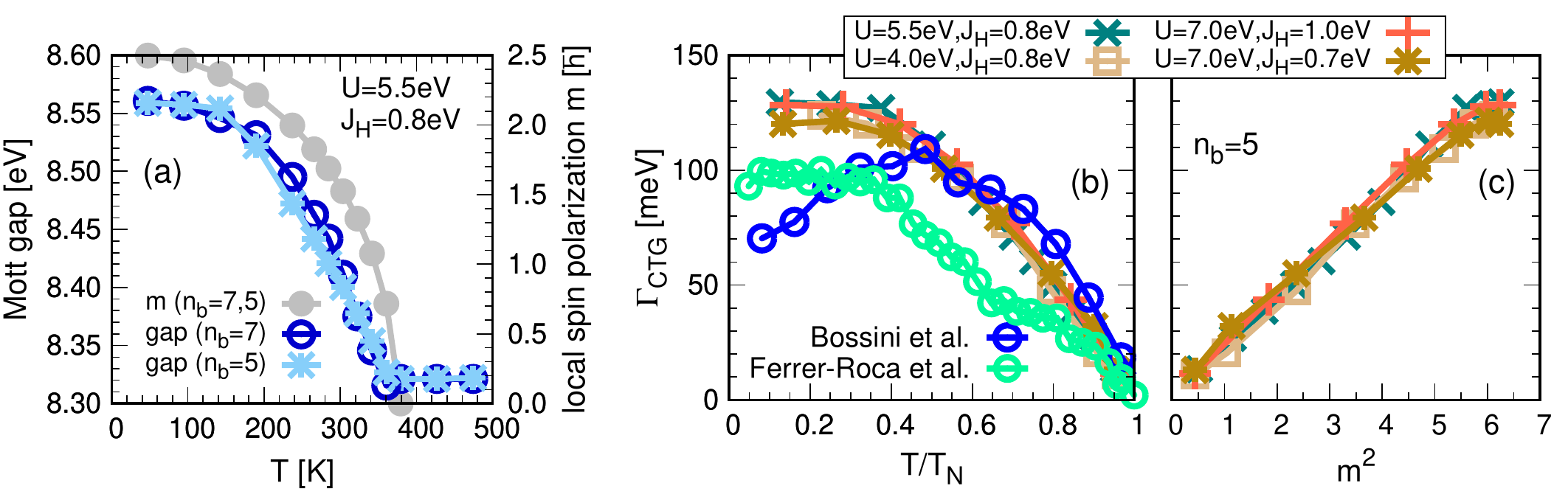}
   \caption{
	 (a) Theoretical results for the Mott gap and the 
	local spin polarization $m$ vs.\ $T$. 
	(b) Theoretical and experimental \cite{Bossini2020,Ferrer-Roca2000} results for the 
	MBS in $\alpha$-MnTe 	as function of temperature.
	(c) MBS  vs.\ the squared spin polarization $m^2$ 
	combining the theoretical results from (a) and (b) for various values 
	 of $U$ and $J\h$ for $n_b=5$.}
   \label{fig:mnte}
   \end{center}
\end{figure*}

The DMFT accurately accounts for the local interactions $U$ and $J\h$. In the limit
of infinite coordination number justifying DMFT, the intersite interactions
$J_n$ are consistently treated by static mean-fields \cite{mulle89a}.
Thus, the intersite magnetic interactions are represented by
\be
\label{eq:MF}
H_\text{MF} =  -\sum_{i}( h_i^{\rm loc} S_i^z + h_i^{\rm iti} s_i^z )
\ee
where the effective magnetic fields 
\bseq
\begin{align}
h_i^{\rm loc} &=2(J_1-3J_2+6J_3-J_4) \langle S_i^z +s_i^z \rangle
\\
h_i^{\rm iti} &=2(J_1-3J_2+6J_3-J_4  ) \langle S_i^z \rangle
\end{align}
\eseq
act on the localized spin ($h_i^{\rm loc}$) and on the itinerant spin ($h_i^{\rm iti}$), respectively.
They need to be determined self-consistently. 
For simplicity, we take the magnetization in $z$-direction although a weak
spin-orbit coupling orients it in $x$-direction \cite{Mu2019,yin19}. But for the spin-isotropic
model studied here this does not matter.  

The resulting Hamiltonian is solved using DMFT starting from
an initial guess  for the self-energy and the local magnetizations $\langle S_{i}^z \rangle$ and 
$\langle s_{i}^z \rangle$. These quantities are updated in each DMFT loop 
until convergence is reached within some tolerance, see App.\ \ref{app:A2}. 
This approach is well justified and goes far beyond the previous two-site calculation \cite{Bossini2020} 
because it properly treats the extended lattice, the dynamics of single charges, and it allows us
to study the temperature dependence. Of course, more sophisticated calculations are conceivable 
in the future to fix 
numerical values to higher accuracy \cite{Fratino2017,Haule2007,Vucicevic2017},
 but our aim here is to elucidate the fundamental physics.

In Fig.\ \ref{fig:mnte}(a) we plot the 
temperature dependence of the Mott gap as obtained with $n_b=5$ and $n_b=7$ bath sites
for $U=5.5$ eV and $J\h=0.8$ eV. 
The agreement of both data sets underlines that the results do not depend
significantly on the number of bath sites. In addition, the local spin polarization
 $m=|\langle S^z_i+s^z_i \rangle|$
is shown, coinciding for $n_b=5$ and $n_b=7$ and indicating a N\'eel temperature  
$T\N\approx 380$ K. This value represents a classical estimate
since the DMFT approach does not capture intersite fluctuations
which are shown \cite{Mu2019} to reduce $T\N$ to $\approx 310$ K 
in accordance with experiment \cite{Walther1967,Szuszkiewicz2005,Kriegner2017}. 
Hence, the effect of the neglected intersite fluctuations on the gap
appears to be about $6$ meV ($\approx 70$ K).

The Mott gap remains almost independent on temperature in the paramagnetic phase 
$T\geq 380$ K in line with our findings in the Hubbard model.
This result supports the assumption in Fig. \ref{fig:bandgap} that 
for $T>T\N$ the Mott gap remains unchanged and the increase of the charge-transfer gap $g$ is essentially due to a smooth 
relative shift of the $p$-band captured by the Varshni fit.
The antiferromagnetic ordering induces a MBS of the Mott gap $\Gamma_{\rm MG}(T)$ 
of approximately $250$ meV as $T\to 0$.
This is the shift 
between the lower $d^-$ and the upper $d^+$ Hubbard bands 
in Fig.\ \ref{fig:bandgap}, i.e., $\Delta d^+ + \Delta d^-$. 
The MBS of the charge-transfer gap $\Gamma_{\rm CTG}(T)$, which is the MBS measured in the 
experiment, is given by $\Delta d^+$.
Since we cannot calculate the individual contributions $\Delta d^\pm$ separately,
we assume that they are shifted symmetrically, i.e., 
$\Delta d^+ = \Delta d^-$ typical for a half-filled Mott insulator.
This implies that the theoretical MBS of the charge-transfer gap
$\Gamma_{\rm CTG}(T)=\Gamma_{\rm MG}(T)/2$ is about 120 meV at its maximum.

Fig.\ \ref{fig:mnte}(b) shows $\Gamma_{\rm CTG}$ as function of $T/T_{\rm N}$. For all four
pairs of $U$ and $J\h$ we find the same N\'eel temperature $T\N\approx 380$ K
which is to be expected since  $T\N$ is determined from the low-energy Heisenberg
model defined by the intersite exchange couplings $J_n$
which we kept fixed; for tables of the explicit parameters used, see
App.\ \ref{app:E}.
But the Mott gap changes significantly from $\approx 7$ eV for 
$U=4.0$ eV and $J\h=0.8$ eV to $\approx 11$ eV for $U=7.0$ eV and $J\h=1.0$ eV.
Remarkably, there is hardly any change in the MBS despite this large change in the Mott gap. 
This corroborates that the essential parameters for the MBS are the exchange
couplings and the hopping elements as indicated by Eq.\ \eqref{eq:fit-HK} above.

We also added the experimental results for the MBS
to Fig.\ \ref{fig:mnte}(b) .
%also includes the experimental results for the MBS. 
The theoretical data agree nicely with the data of Bossini et al.\ \cite{Bossini2020} for 
$T \gtrapprox 0.5T\N$. It is mentioned by Ferrer-Roca et al. \cite{Ferrer-Roca2000} that their
data probably underestimates the MBS between $T\approx 0.45 T\N$ and $T\approx 0.65 T\N$.
The results of Bossini et al.\ deviate from theory below $T \approx 0.5T\N$ 
where the experimental data turn down in contrast to expectation and the data set from Ref.\  \cite{Ferrer-Roca2000}. The 
deviating downturn is likely due to experimental reasons, e.g., sample quality
and/or stability of the experimental conditions. 
In fact, due to the saturation of the spin polarization at low temperatures we expect the MBS also to 
saturate as is found by Ferrer-Roca et al. for $T<0.4T_{\rm N}$.
We emphasize that the very good
agreement between experiment and theory in Fig.\ \ref{fig:mnte}(b)
is achieved using generic parameters 
 from literature for the Hubbard-Kondo model \emph{without} any fine-tuning
in contrast to the approach in Ref.\ \cite{Ando1992}.
This provides strong evidence that the observed blue-shift upon ordering
is the generic MBS of the advocated Hubbard-Kondo lattice model. 

Finally, Fig.\ \ref{fig:mnte}(c) combines the calculated  $\Gamma_{\rm CTG}(T)$
and the local spin polarization $m(T)$ 
eliminating the temperature. The value of the gap at $T\N$ is fixed 
such that the MBS vanishes for $m^2\rightarrow 0$. The figure clearly shows that there is an almost
linear relation between the squared local spin polarization $m^2$ and the MBS, $\Gamma_{\rm CTG}(T)\propto m^2(T)$,
for various parameter combinations of the local interactions $U$ and $J\h$. This behavior is in line
with previous experimental findings \cite{Ferrer-Roca2000} 
and underlines that the MBS is a robust effect not depending 
on details. Ref.\ \cite{Ando1992} also finds a MBS $\propto m^2$, but the 
computation strongly depends on the chosen parameters since it originates 
from scenario (i), 
in which the magnetic order affects the itinerant electrons only indirectly.

\begin{figure}[t]
   \begin{center}
   \includegraphics[height=0.8\columnwidth,angle=-90]{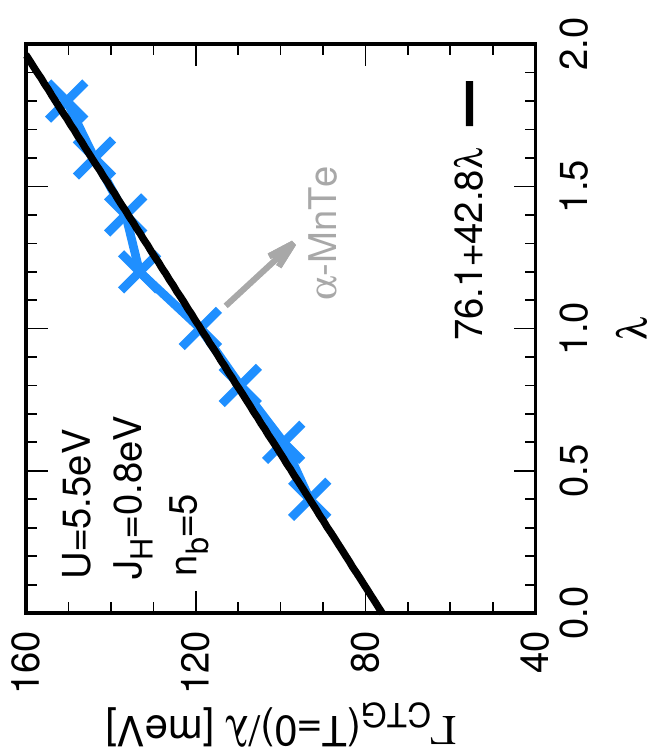}
   \caption{MBS at $T=0$ as function of the scaling parameter of the hopping $\lambda$
	 as given in Eqs.\ \eqref{eq:scaling}. As in Fig.\ \ref{fig:mbs},
	the MBS offset on the $y$-axis is proportional to the hopping stemming from
	the double-exchange mechanism while the slope results from the contribution
	to the MBS proportional to the magnetic exchange couplings. Note that for $\alpha$-MnTe ($\lambda=1$) the hopping
	contribution equals roughly twice the exchange contribution.}
   \label{fig:scaling-MnTe}
   \end{center}
\end{figure}

The MBS in Hubbard-Kondo models is governed by a contribution from the hopping and a 
contribution from the intersite exchange interaction, see Eq.\ \eqref{eq:fit-HK}. 
We investigate this point for $\alpha$-MnTe as well by means of a plot
analogous to Fig.\ \ref{fig:mbs}. In view of the numerous parameters
relevant for $\alpha$-MnTe (four hopping elements, four exchange couplings, the Hubbard 
interaction and the Hund coupling) 
a variation of individual parameters appears to be not practical.
Hence, we resort to a uniform scaling by a parameter $\lambda$ according to
\bseq
\label{eq:scaling}
\begin{align}
t_n &\to t_n(\lambda)= \lambda t_n
\\
\Rightarrow \quad J_n &\to J_n(\lambda)= \lambda^2 J_n
\end{align}
\eseq
leaving the local interactions $U$ and $J_H$ unchanged. 
The plot $\Gamma_\text{CTG}(T=0)/\lambda$ vs.\ $\lambda$ for $U=5.5$ eV and $J_{\rm H}=0.8$ eV in Fig.\ \ref{fig:scaling-MnTe} 
recreates the same kind of analytical dependence as Fig.\ \ref{fig:mbs}.
Note that for $\lambda<2$ we have $\lambda t_n/\Delta<0.02$ which corresponds to the deep Mott regime.
As expected, we find the same qualitative behavior as in Fig.\ \ref{fig:mbs} described very well by a linear fit.
The offset at $\lambda=0$ is the contribution proportional to the hoppings due to the
rescaled $\Gamma_\text{CTG}(0)/\lambda$ plotted. In the same way, the slope results
from contributions proportional to the magnetic exchange couplings $J_n(\lambda)\propto \lambda^2$.
The plot in Fig.\ \ref{fig:scaling-MnTe} again allows us to separate the two contributions
as the plot in Fig.\ \ref{fig:mbs} did for the Hubbard-Kondo model.
The nice linear behavior suggests a hopping contribution of about $76$ meV and an exchange contribution 
of about $43$ meV to the MBS in $\alpha$-MnTe. This large hopping contribution emphasizes the 
important role of the double-exchange mechanism on the MBS in systems with localized spins.

\section{Conclusions}
\label{sec:conclusio}

We established that the MBS in the Mott regime of the 3D Hubbard model stems from 
the magnetic exchange coupling. While the decrease of the MBS with increasing interaction $U$ 
had been observed before \cite{Fratino2017} its proportionality to the magnetic exchange
$J=4t^2/U$ is a new finding, which sets the energy scale for further contributions.

Our key result relates to systems which involve localized spins in addition to itinerant electrons. 
For the Hubbard-Kondo model with a ferromagnetic Kondo coupling
we showed that there are two contributions to the MBS: one similar 
to the MBS in the Hubbard model which is proportional to the magnetic exchange, 
and another which is proportional to the hopping. The latter stems from
the double-exchange mechanism which reduces the effective hopping between sites
with antiparallel spin ordering.

This finding opens up a route to applications of the MBS
 since a plethora of heavily investigated systems consists of itinerant electrons 
and localized spins, for instance the manganites. Exemplarily, we elucidated the 
origin of the experimentally established MBS in $\alpha$-MnTe which is a promising
candidate for applications with AF order at room temperature.
We developed an extended Hubbard-Kondo lattice model for $\alpha$-MnTe.
The MBS found in this model is in overall excellent agreement with the experimental findings 
for $\alpha$-MnTe.

Strong MBSs in magnetic semiconductors can play a major role
in spin-to-charge conversion on the femtosecond time scale, 
which is the characteristic time scale of the hopping and the intersite exchange interactions.
Recent progress in the manipulation of magnons in antiferromagnets on
ultrafast time scales \cite{Bossini2016,Bossini2018} add to the relevance of 
a comprehensive understanding of the coupling of spin and charge dynamics
\cite{Gillmeister2020}. A major outlook of our work consists in 
exploring the role of  dimensionality of a Mott system on the MBS.
This is highly relevant in view of both the massive present research activity 
on 2D materials and the widely explored 
properties of low-dimensional magnetic semiconductors.
Hence, the demonstrated MBS paves a promising
route for future research, both fundamental and applied.

\acknowledgments

This study was funded by the German Research Foundation (DFG) and the Russian Foundation 
for Basic Research in the International 
Collaborative Research Centre TRR 160 (Project B8) and by the DFG grant BO 5074/1-1.

% \bibliography{references,references00,references10,references20,references30}

\appendix

\section{Theoretical Approach}
\label{app:A}

\subsection{General remarks}

We use dynamic mean-field theory (DMFT) \cite{Georges1996} which
is an established approach for strong local interactions and 
large coordination number. 
The frequency dependent self-energy allows us to describe paramagnetic 
Mott insulators, not accessible by static mean-field theories.
We use the real-space DMFT (RDMFT) method \cite{Potthoff1999,Song2008,Snoek2008} 
as implemented by one of us
\cite{Hafez-Torbati2018} and applied successfully to various models 
\cite{Hafez-Torbati2019,Hafez-Torbati2020,Ebrahimkhas2021}.
We note that for the bulk properties it is not necessary to use the real-space extension 
of DMFT. But in view of future analysis of the spatial dependence in thin films as 
in Ref. \cite{Bossini2020} we opt for RDMFT for comparability.

Exact diagonalization (ED) is employed as impurity solver \cite{Caffarel1994} 
providing direct access to dynamics at real frequencies and the quantum mechanical 
treatment of localized spins going beyond previous classical approximations based on quantum 
Monte Carlo solver \cite{Furukawa1994,Calderon1998,Held2000}.
The local spectral function $A(\omega)$ results from 
the imaginary part of the local Green's function, 
averaged over both sublattice sites.
We compute the Mott gap from the positions of the peaks in the spectral function.
Although the spectral function for finite number of bath sites $n_b$ consists of a series of 
sharp peaks approximating the continuous function, the Mott gap is found accurate and 
is used to benchmark the results of other methods \cite{Wang2009}.
We use the chemical potential $\mu$ to satisfy the  condition of half-filling.
The lattice system is approximated by clusters of
$L\times L\times L$ sites with $L=10$.  
We checked for selected temperatures close to the transition
temperature that the results remain the same for $L=20$.

\subsection{Dynamical mean-field theory of the Hubbard-Kondo model}
\label{app:A2}

After the mean-field decoupling of the intersite magnetic interactions 
shown in \eqref{eq:MF} the Hamiltonian from \eqref{eq:HKM_main} reads
\bearr
\label{eq:HKM}
H=&-&\sum_{i,j}\sum_{\sigma} t_{j,i}^\vpdag 
\left( c^{\dagg}_{j,\sigma}c^{\vpdag}_{i,\sigma}+{\rm H.c.}\right) 
- \sum_i (h_i^{\rm iti} s_i^z+\mu n_i^{\vpdag})
\nn \\
&+& U \sum_{i} n^{\vpdag}_{i,\downarrow} n^{\vpdag}_{i,\uparrow} 
\!-\! 2J^\vpdag_{\!H}\sum_{i} \vec{S}_i \cdot \vec{s}_i
-\sum_{i} h_i^{\rm loc} S_i^z 
\eearr
with the effective magnetic fields 
\bseq
\label{eq:fields}
\begin{align}
h_i^{\rm loc}&=2(J_1-3J_2+6J_3-J_4  ) \langle S_i^z +s_i^z \rangle, \\
h_i^{\rm iti}&=2(J_1-3J_2+6J_3-J_4  ) \langle S_i^z \rangle 
\end{align}
\eseq
acting on the localized spin $\vec{S}_i$ and on the spin of the itinerant 
electrons $\vec{s}_i$, respectively, at the lattice site $i$. 
We added a chemical potential term $\mu$ to the Hamiltonian \eqref{eq:HKM} 
to control the electron density 
$n^{\vpdag}_i:=n^{\vpdag}_{i,\downarrow}+n^{\vpdag}_{i,\uparrow}$
in the system keeping it at half filling. In the derivation of Eq.\ \eqref{eq:fields} 
we consider ferromagnetic order within the triangular layers and antiferromagnetic order between 
them. For simplicity, the magnetic order is taken to be in the $\hat{z}$ direction in spin space,
but the choice of direction does not matter since we consider a fully spin isotropic model.
The treatment of the weak anisotropy stemming from a spin-orbit coupling \cite{Mu2019,yin19} 
is left to future research.
The effective magnetic fields in Eq.\ \eqref{eq:fields} depend
on the local spin polarizations $\langle S_i^z \rangle$ and $\langle s_i^z \rangle$ 
and need to be determined self-consistently in the course of the iterations of the RDMFT. 

Essentially, we use the RDMFT implementation of Ref.\ \cite{Hafez-Torbati2018} for SU($2$) systems with a 
generalization of the Anderson impurity model to an Anderson-Kondo impurity model which includes 
the additional local degrees of freedom, here the localized spin in  Eq.\ \eqref{eq:HKM}. 
Note that we treat the spin fully quantum mechanically.
We also updated the implementation of Ref.\ \cite{Hafez-Torbati2018} such that some local expectation
values are computed during the RDMFT loop so that the mean-fields
can be modified iteratively. In the case 
of the Hamiltonian  \eqref{eq:HKM} these local expectation values
 are  the spin polarizations $\langle s_i^z \rangle$ and $\langle S_i^z\rangle$, 
needed for the calculations of the effective magnetic fields Eq.\ \eqref{eq:fields}.
We stress that the local Green's function, the self-energy, 
and the dynamical Weiss field are all diagonal in spin space as the 
Hamiltonian Eq.\ \eqref{eq:HKM} is diagonal in  $S^z$.
This simplifies the general formalism of Ref.\ \cite{Hafez-Torbati2018}.

The terms in the first line of Eq.\ \eqref{eq:HKM} describe the 
non-interacting parts of the itinerant electrons 
from which the non-interacting lattice Green's function is constructed.
The second line in  Eq.\ \eqref{eq:HKM} contains the Hubbard interaction between 
the itinerant electrons, the Hund coupling between the
spin of the itinerant electron and the  localized spin $S=2$, 
and the effective magnetic field at the localized spin. They
 enter the calculation in the local impurity problem.
The RDMFT loop starts with an initial guess for the self-energy matrix 
$\bs{\Sigma}(i\omega_n)$ and the local spin polarizations
$\langle s_i^z \rangle$ and $\langle S_i^z\rangle$. The real-space lattice Green's 
function is calculated according to Dyson's equation
\be
\label{eq:green}
{\bs G}(i\omega_n)
= \left[ i\omega_n \mathds{1} -{\bs H}_0- {\bs \Sigma}(i\omega_n) \right]^{-1},
\ee
where ${\bs H}_0$ is the matrix representation of the non-interacting terms in the first line
of Eq.\ \eqref{eq:HKM}. To address the local problem at the lattice site $i$ 
we use the Anderson-Kondo impurity model \cite{Peters2006}
\bearr
\label{eq:AKM}
&&H_i^{\vpdag}=-\mu n^{\vpdag}_i -h_i^{\rm iti} s_i^z 
+ U n^{\vpdag}_{i,\downarrow} n^{\vpdag}_{i,\uparrow}
-h_i^{\rm loc} S_i^z - 2J^\vpdag_{\!H} \vec{S}_i \cdot \vec{s}_i 
\nn \\
&+&\sum_{\ell=1}^{n_b} \sum_{\sigma} \epsilon^{i}_{\ell} 
a^{\dagg}_{\ell,\sigma} a^{\vpdag}_{\ell,\sigma}
+\sum_{\ell=1}^{n_b} \sum_{\sigma} \left(a^{\dagg}_{\ell,\sigma} 
V^i_{\ell,\sigma} c^{\vpdag}_{i,\sigma}+{\rm H.c.} \right)
\eearr
where $a^{\dagg}_{\ell,\sigma}$ and $a^{\vpdag}_{\ell,\sigma}$ are the fermionic creation and 
annihilation operators at the bath site $\ell$ with the spin $\sigma=\uparrow,\downarrow$.
The bath sites in Eq.\ \eqref{eq:AKM} approximate the effect of the surrounding sites 
in the lattice \cite{Georges1996}.
The bath parameters $\epsilon^{i}_{\ell}$ and $V^i_{\ell,\sigma}$ are determined by  
fitting the dynamical Weiss field \cite{Caffarel1994,Hafez-Torbati2018}. The self-energy 
as well as the local spin polarizations $\langle s_i^z \rangle$ and 
$\langle S_i^z\rangle$ are calculated 
using ED of the Anderson-Kondo impurity model \eqref{eq:AKM}.
These quantities are employed for the next RDMFT iteration loop. 

Since the model is symmetric with respect to a combined swap of the sublattice
and the spin orientations, we only need to set up the impurity 
model \eqref{eq:AKM} for one representative site. In this sense, the lattice
solutions are homogeneous. Hence, one does not need to 
fully invert the matrix in Eq.\ \eqref{eq:green} because only the two columns 
for the two spin orientations at the representative site are needed 
\cite{Hafez-Torbati2018}.
This enables us to treat very large system sizes in Eq.\ \eqref{eq:green} 
 so that finite-size corrections are completely negligible.

\section{Internal energy and magnetic blue-shift}
\label{app:D}

The internal energy of a general interacting fermionic system described by the Hamiltonian
\be
H=H_0+W=\sum_{i,j} h_{i,j}^\vpdag c^\dag_i c^\vpdag_j 
+\frac{1}{2} \sum_{i,j,k,l} W_{i,j,k,l}^\vpdag c^\dag_i c^\dag_j c^\vpdag_k c^\vpdag_l \ ,
\label{eq:H}
\ee
can be expressed as \cite{Fetter2012}
\be 
E := 
\langle H \rangle = \frac{1}{2} \sum_{i,j} \int_{-\infty}^{+\infty}
{\rm d} \omega A^\vpdag_{i,j}(\omega)f(\omega) [\omega \delta_{i,j}+h_{i,j} ] \ ,
\label{eq:E}
\ee
where $i$ and $j$ specify single-particle quantum numbers, $A^\vpdag_{i,j}(\omega)$ is the 
spectral function of the single-particle 
Green's function, and $f(\omega)$ is Fermi's occupation function. 
Eq.\ \eqref{eq:E} shows that the internal energy can be determined 
solely from the single-particle spectral function.
The first contribution in Eq.\ \eqref{eq:E} describes $\langle H_0^\vpdag\rangle /2 +\langle W \rangle$  
while the second contribution equals half the kinetic energy, $\langle H_0^\vpdag\rangle/2$.

The Mott gap separating the lower and the upper Hubbard bands is typically much larger than 
the N\'eel temperature $T_{\rm N}$. For temperatures $T \lessapprox T_{\rm N}$ this essentially
restricts the integration in Eq.\ \eqref{eq:E} to only the lower Hubbard band (LHB). 
Then, the first contribution in Eq.\ \eqref{eq:E}  can be simplified to 
\be
\varepsilon_1=\frac{E_1}{N} = \int_{\rm LHB}
 \omega A(\omega) {\rm d} \omega \ ,
 \label{eq:E1}
\ee
where $N$ is the number of lattice sites and we used the translational symmetry of the spin-averaged local 
spectral function $A(\omega)$, which we plotted in Fig.\ \ref{fig:sp:hm:T}. 

\begin{figure}[t]
   \begin{center}
   \includegraphics[width=0.85\columnwidth]{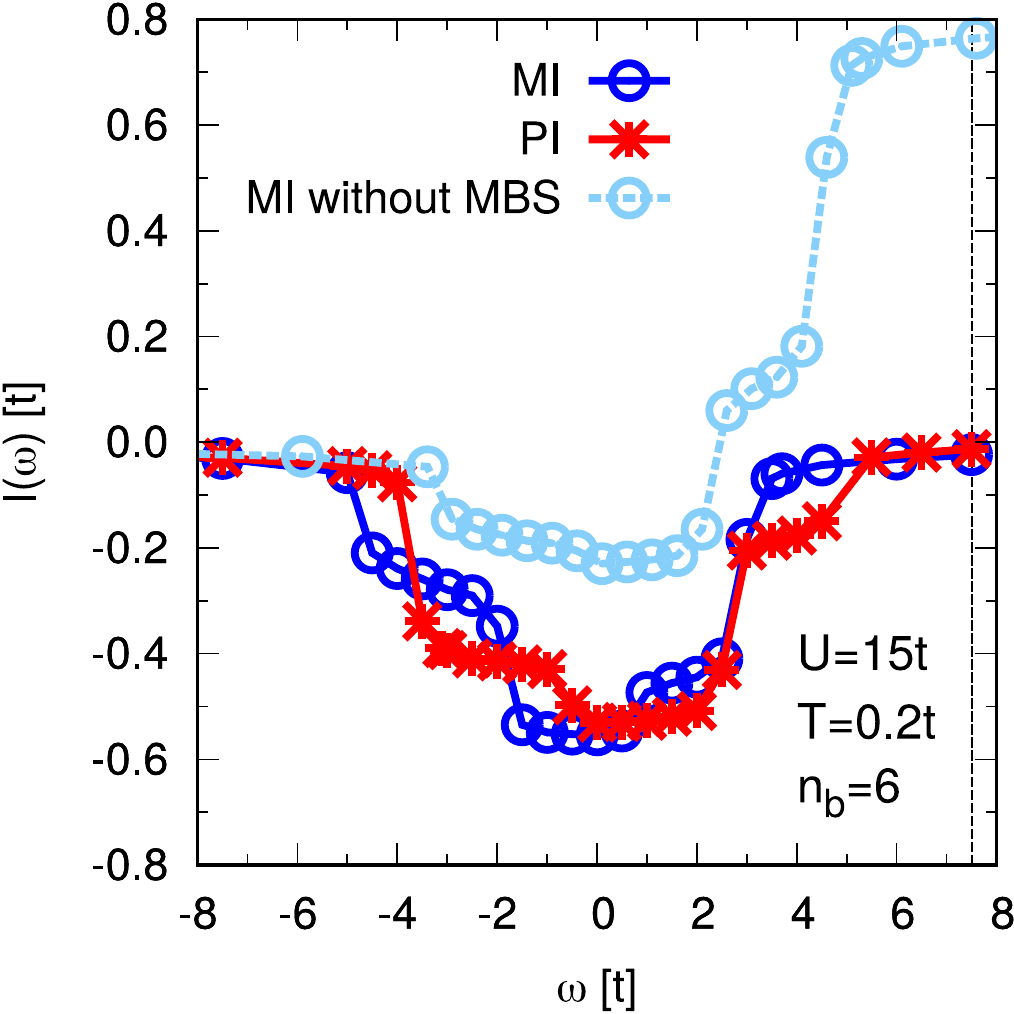}
   \caption{The partial energy $I(\omega)$ defined in Eq.\ \eqref{eq:pE} versus frequency $\omega$ changing over the lower Hubbard band 
   for the 3D Hubbard model for $U=15t$ and $T=0.2t$ 
	in the MI (dark blue) and in the PI (red)
   as well as in the MI without MBS (light blue). 
   The vertical dashed line at $U/2=7.5t$ shows the Fermi energy.
   The number of bath sites is $n_b=6$ in the ED impurity solver.}
   \label{fig:pE}
   \end{center}
\end{figure}

In order to see how the energy in Eq.\ \eqref{eq:E1} is distributed over  frequency 
we consider the partial energy
\be
I(\omega)= \int_{-\infty}^{\omega}  \omega' A(\omega') {\rm d} \omega' \ ,
 \label{eq:pE}
\ee
which equals $\varepsilon_1$ if $\omega$ is large enough to cover the whole LHB.
In Fig.\ \ref{fig:pE} we plot $I(\omega)$ for the 3D Hubbard model at $U=15t$ and $T=0.2t$ 
in the MI and in the PI as well as in the MI without any MBS
of the local spectral function. The results are obtained using  $n_b=6$  bath sites in the 
ED impurity solver. Fig.\ \ref{fig:pE} clearly shows that upon entering the magnetically ordered phase 
from the paramagnetic phase a redistribution of the weight within the spectral function occurs which 
leads to a large increase in the internal energy \emph{if} it is not compensated by a MBS. 
Such a redistribution has been observed also in Ref.\ \cite{Sangiovanni2006}, 
both experimentally and theoretically.

\begin{figure}[t]
   \begin{center}
   \includegraphics[width=0.8\columnwidth,angle=-90]{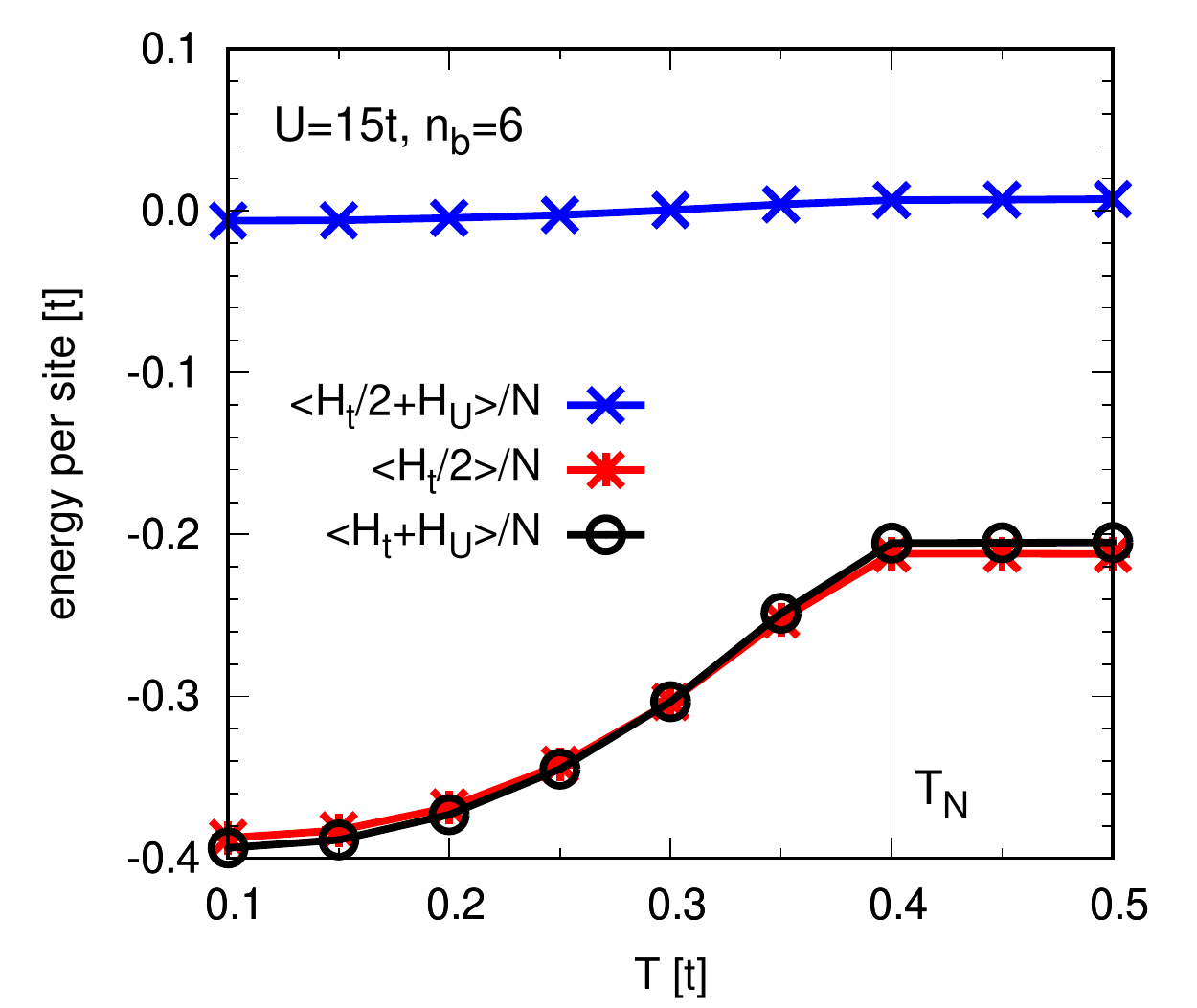}
   \caption{The internal energy $\langle H_t+H_U \rangle$ and the individual contributions 
	$\langle H_t/2+H_U \rangle$ and 
   $\langle H_t/2 \rangle$ of the Hubbard model $H=H_t+H_U$ per lattice site versus the 
	temperature $T$ for the Hubbard interaction $U=15t$ 
   and $n_b=6$  bath sites in the ED impurity solver.}
   \label{fig:iE}
   \end{center}
\end{figure}

For the Hubbard model $H=H_t+H_U$ with the nearest-neighbor hopping term $H_t$ and the Hubbard interaction $H_U$ we plot
the internal energy $\langle H_t+H_U \rangle$ per lattice site in Fig.\ \ref{fig:iE} for $U=15t$ and $n_b=6$. We include
also $\langle H_t/2+H_U \rangle$ and $\langle H_t/2 \rangle$ corresponding to the first and the second 
contribution in Eq.\ \eqref{eq:E}, respectively. We see that $\langle H_t/2+H_U \rangle$ remains close to zero and the reduction of
the internal energy below $T_{\rm N}$ is mainly due to $\langle H_t/2 \rangle$. Without MBS the contribution 
$\langle H_t/2+H_U \rangle$ would increase substantially in the MI phase, see Fig.\ \ref{fig:pE}.
This shows that the MBS is crucial to achieve a decrease of the internal energy which is 
the prerequisite for the phase transition into the ordered phase to occur.

\section{Model parameters for $\alpha$-${\rm \bf MnTe}$}
\label{app:E}

We fixed the intersite exchange interactions in $\alpha$-MnTe 
according to the value from Ref.\ 
\cite{Mu2019}: $J_1=3.072$ meV, $J_2=0.0272$ meV, $J_3=0.4$ meV, and $J_4=0.16$ meV.
The hopping parameters corresponding 
to the different sets of the Hubbard interaction $U$ and the Hund coupling 
$J_{\rm H}$ are calculated from the relation
\be
\label{eq:t-J}
J_n=4t_n^2/\Delta
\ee
with $\Delta=U+4J_{\rm H}$. They are given in Table \ref{tab:para1} for future 
convenient use.

Expressing energies in units of $t_1$, as used in the DMFT calculations, 
one has the hopping parameters 
\be
t_n=t_1\sqrt{J_n/J_1}
\ee
independent of the choice of $U$ and $J_{\rm H}$. 
They are given by $t_2=0.0941t_1$, $t_3=0.3608t_1$, and $t_4=0.2282t_1$. 
But the intersite exchange couplings $J_n$ 
in units of $t_1$ depend on $U$ and $J_{\rm H}$. 
Table \ref{tab:para2} provides $J_n$, $U$, and $J_{\rm H}$ in units 
of $t_1$ corresponding to the different parameter sets used in Table \ref{tab:para1}.

\begin{table}[ht]
\begin{center}
    \begin{tabular}{ l | l | l | l | l }
    $(U,J_{\rm H})$  [eV] & $t_1$ [meV] & $t_2$ [meV] & $t_3$ [meV] & $t_4$ [meV] \\ \hline
    $(7.0,1.0)$ & $91.91$ & $8.649$ & $33.16$ & $20.97$ \\ \hline
    $(7.0,0.7)$ & $86.75$ & $8.163$ & $31.30$ & $19.80$ \\ \hline
    $(5.5,0.8)$ & $81.74$ & $7.692$ & $29.49$ & $18.65$ \\ \hline
    $(4.0,0.8)$ & $74.36$ & $6.997$ & $26.83$ & $16.97$ \\ 
    \end{tabular}
    \caption{The hopping parameters $t_n$ in $\alpha$-MnTe according to Eq.\ 
		\eqref{eq:t-J} for the various sets of 
     Hubbard interaction $U$ and Hund coupling $J_{\rm H}$. The hopping parameters are 
    in units of meV and $U$ and $J_{\rm H}$ in units of eV. Their subscripts 
		refer to the numbers in Fig.\ \ref{fig:HKM}(b).}
    \label{tab:para1}
\end{center}
\end{table}

\begin{table}[hb]
\begin{center}
    \begin{tabular}{ l | l | l | l | l }
    $(U,J_{\rm H})$ & $J_1 [10^{-2}]$ & $J_2 [10^{-2}]$ & $J_3 [10^{-2}]$ & $J_4 [10^{-2}]$ \\ \hline
    $(76.16,10.88)$ & $3.342$ & $0.02959$ & $0.4352$ & $0.1741$ \\ \hline
    $(80.69,8.07)$ & $3.541$ & $0.03135$ & $0.4611$ & $0.1844$ \\ \hline
    $(67.29,9.787)$ & $3.758$ & $0.03328$ & $0.4894$ & $0.1957$ \\ \hline
    $(53.79,10.76)$ & $4.131$ & $0.03658$ & $0.5379$ & $0.2152$ \\ 
    \end{tabular}
    \caption{The intersite exchange couplings $J_n$, the Hubbard interaction $U$, 
    and the Hund coupling $J_{\rm H}$ in units of $t_1$. 
		The rows correspond to the different parameter 
    sets used in Table \ref{tab:para1}.}
    \label{tab:para2}
\end{center}
\end{table}

\end{document}